\documentclass[
aps,%
12pt,%
final,%
notitlepage,%
oneside,%
onecolumn,%
nobibnotes,%
nofootinbib,%
superscriptaddress,%
noshowpacs,%
centertags]%
{revtex4}

\newcommand{\tw}{0.95\textwidth}

\newcommand{\htw}{0.47\textwidth}

\newcommand{\fig}[1]{Fig.\,\ref{fig:#1}}

\begin{document}
\selectlanguage{english}

\title{Search for Astrophysical Nanosecond Optical Transients with TAIGA-HiSCORE Array}

\author{\firstname{A.~D.}~\surname{Panov}}
\email{panov@dec1.sinp.msu.ru}
\affiliation{%
Lomonosov Moscow State University Skobeltsyn Institute of Nuclear Physics (MSU SINP),
Leninskie gory 1(2), GSP-1, Moscow, 119991, Russia.
}%
\author{\firstname{I.~I.}~\surname{Astapov}}
\affiliation{%
National Research Nuclear University MEPhI (Moscow Engineering Physics Institute),
Kashirskoe highway 31, Moscow, 115409, Russia.
}%
\author{\firstname{A.~K.}~\surname{Awad}}
\affiliation{%
Institute of experimental physics of Hamburg University, Luruper Chaussee 149, 22761
Hamburg, Germany.
}%
\author{\firstname{G.~M.}~\surname{Beskin}}
\affiliation{%
Special Astrophysical Observatory, Nizhnij Arkhyz, Zelenchukskiy region, Karachai-Cherkessian Republic, 369167, Russia.
}%
\author{\firstname{P.~A.}~\surname{Bezyazeekov}}
\affiliation{%
Institute of Applied Physics, Irkutsk State University (API ISU), Gagarin Blvd. 20, Irkutsk,
664003, Russia.
}%
\author{\firstname{M.}~\surname{Blank}}
\affiliation{%
Institute of experimental physics of Hamburg University, Luruper Chaussee 149, 22761
Hamburg, Germany.
}%
\author{\firstname{E.~А.}~\surname{Bonvech}}
\affiliation{%
Lomonosov Moscow State University Skobeltsyn Institute of Nuclear Physics (MSU SINP),
Leninskie gory 1(2), GSP-1, Moscow, 119991, Russia.
}%
\author{\firstname{A.~N.}~\surname{Borodin}}
\affiliation{%
Joint Institute for Nuclear Research, Joliot-Curie 6, Dubna, Moscow Region, 141980, Russia.
}%
\author{\firstname{M.}~\surname{Br\"uckner}}
\affiliation{%
Deutsches Elektronen-Synchrotron DESY, 15738 Zeuthen, Germany.
}%
\author{\firstname{N.~M.}~\surname{Budnev}}
\affiliation{%
Institute of Applied Physics, Irkutsk State University (API ISU), Gagarin Blvd. 20, Irkutsk,
664003, Russia.
}%
\author{\firstname{A.~V.}~\surname{Bulan}}
\affiliation{%
Lomonosov Moscow State University Skobeltsyn Institute of Nuclear Physics (MSU SINP),
Leninskie gory 1(2), GSP-1, Moscow, 119991, Russia.
}%
\author{\firstname{D.~V.}~\surname{Chernov}}
\affiliation{%
Lomonosov Moscow State University Skobeltsyn Institute of Nuclear Physics (MSU SINP),
Leninskie gory 1(2), GSP-1, Moscow, 119991, Russia.
}%
\author{\firstname{А.}~\surname{Chiavassa}}
\affiliation{%
Physics Department of the University of Torino and the National Institute of Nuclear
Physics INFN, 10125 Torino, Italy.
}%
\author{\firstname{A.~N.}~\surname{Dyachok}}
\affiliation{%
Institute of Applied Physics, Irkutsk State University (API ISU), Gagarin Blvd. 20, Irkutsk,
664003, Russia.
}%
\author{\firstname{A.~R.}~\surname{Gafarov}}
\affiliation{%
Institute of Applied Physics, Irkutsk State University (API ISU), Gagarin Blvd. 20, Irkutsk,
664003, Russia.
}%
\author{\firstname{A.~Yu.}~\surname{Garmash}}
\affiliation{%
Novosibirsk State University, Pirogova 1, Novosibirsk, 630090, Russia.
}%
\affiliation{%
Budker Institute of Nuclear Physics of the Siberian Branch of the Russian Academy of
Sciences, Lavrentyev Prosp. 11, Novosibirsk, 630090, Russia.
}%
\author{\firstname{V.~M.}~\surname{Grebenyuk}}
\affiliation{%
Joint Institute for Nuclear Research, Joliot-Curie 6, Dubna, Moscow Region, 141980, Russia.
}%
\author{\firstname{O.~A.}~\surname{Gress}}
\affiliation{%
Institute of Applied Physics, Irkutsk State University (API ISU), Gagarin Blvd. 20, Irkutsk,
664003, Russia.
}%
\author{\firstname{T.~I.}~\surname{Gress}}
\affiliation{%
Institute of Applied Physics, Irkutsk State University (API ISU), Gagarin Blvd. 20, Irkutsk,
664003, Russia.
}%
\author{\firstname{A.~A.}~\surname{Grinyuk}}
\affiliation{%
Joint Institute for Nuclear Research, Joliot-Curie 6, Dubna, Moscow Region, 141980, Russia.
}%
\author{\firstname{O.~G.}~\surname{Grishin}}
\affiliation{%
Institute of Applied Physics, Irkutsk State University (API ISU), Gagarin Blvd. 20, Irkutsk,
664003, Russia.
}%
\author{\firstname{D.}~\surname{Horns}}
\affiliation{%
Institute of experimental physics of Hamburg University, Luruper Chaussee 149, 22761
Hamburg, Germany.
}%
\author{\firstname{A.~L.}~\surname{Ivanova}}
\affiliation{%
Budker Institute of Nuclear Physics of the Siberian Branch of the Russian Academy of
Sciences, Lavrentyev Prosp. 11, Novosibirsk, 630090, Russia.
}%
\affiliation{%
Institute of Applied Physics, Irkutsk State University (API ISU), Gagarin Blvd. 20, Irkutsk,
664003, Russia.
}%
\author{\firstname{N.~N.}~\surname{Kalmykov}}
\affiliation{%
Lomonosov Moscow State University Skobeltsyn Institute of Nuclear Physics (MSU SINP),
Leninskie gory 1(2), GSP-1, Moscow, 119991, Russia.
}%
\author{\firstname{V.~V.}~\surname{Kindin}}
\affiliation{%
Institute of Applied Physics, Irkutsk State University (API ISU), Gagarin Blvd. 20, Irkutsk,
664003, Russia.
}%
\author{\firstname{S.~N.}~\surname{Kiryuhin}}
\affiliation{%
Institute of Applied Physics, Irkutsk State University (API ISU), Gagarin Blvd. 20, Irkutsk,
664003, Russia.
}%
\author{\firstname{R.~P.}~\surname{Kokoulin}}
\affiliation{%
National Research Nuclear University MEPhI (Moscow Engineering Physics Institute),
Kashirskoe highway 31, Moscow, 115409, Russia.
}%
\author{\firstname{K.~G.}~\surname{Kompaniets}}
\affiliation{%
Institute of Applied Physics, Irkutsk State University (API ISU), Gagarin Blvd. 20, Irkutsk,
664003, Russia.
}%
\author{\firstname{E.~E.}~\surname{Korosteleva}}
\affiliation{%
Lomonosov Moscow State University Skobeltsyn Institute of Nuclear Physics (MSU SINP),
Leninskie gory 1(2), GSP-1, Moscow, 119991, Russia.
}%
\author{\firstname{V.~A.}~\surname{Kozhin}}
\affiliation{%
Lomonosov Moscow State University Skobeltsyn Institute of Nuclear Physics (MSU SINP),
Leninskie gory 1(2), GSP-1, Moscow, 119991, Russia.
}%
\author{\firstname{E.~A.}~\surname{Kravchenko}}
\affiliation{%
Novosibirsk State University, Pirogova 1, Novosibirsk, 630090, Russia.
}%
\affiliation{%
Budker Institute of Nuclear Physics of the Siberian Branch of the Russian Academy of
Sciences, Lavrentyev Prosp. 11, Novosibirsk, 630090, Russia.
}%
\author{\firstname{A.~A.}~\surname{Krivopalova}}
\noaffiliation%
\author{\firstname{L.~A.}~\surname{Kuzmichev}}
\email{kuz@dec1.sinp.msu.ru}
\affiliation{%
Lomonosov Moscow State University Skobeltsyn Institute of Nuclear Physics (MSU SINP),
Leninskie gory 1(2), GSP-1, Moscow, 119991, Russia.
}%
\author{\firstname{A.~P.}~\surname{Kryukov}}
\affiliation{%
Lomonosov Moscow State University Skobeltsyn Institute of Nuclear Physics (MSU SINP),
Leninskie gory 1(2), GSP-1, Moscow, 119991, Russia.
}%
\author{\firstname{A.~A.}~\surname{Lagutin}}
\affiliation{%
Altai State University, Lenina 61, Barnaul, 656049, Russia.
}%
\author{\firstname{M.~V.}~\surname{Lavrova}}
\affiliation{%
Joint Institute for Nuclear Research, Joliot-Curie 6, Dubna, Moscow Region, 141980, Russia.
}%
\author{\firstname{Yu.}~\surname{Lemeshev}}
\affiliation{%
Institute of Applied Physics, Irkutsk State University (API ISU), Gagarin Blvd. 20, Irkutsk,
664003, Russia.
}%
\author{\firstname{B.~K.}~\surname{Lubsandorzhiev}}
\affiliation{%
Institute for Nuclear Research of the Russian Academy of Sciences, 60th October
Anniversary 7a, 117312, Moscow, Russia.
}%
\author{\firstname{N.~B.}~\surname{Lubsandorzhiev}}
\affiliation{%
Lomonosov Moscow State University Skobeltsyn Institute of Nuclear Physics (MSU SINP),
Leninskie gory 1(2), GSP-1, Moscow, 119991, Russia.
}%
\author{\firstname{A.~D.}~\surname{Lukanov}}
\affiliation{%
Institute for Nuclear Research of the Russian Academy of Sciences, 60th October
Anniversary 7a, 117312, Moscow, Russia.
}%
\author{\firstname{R.~R.}~\surname{Mirgazov}}
\affiliation{%
Institute of Applied Physics, Irkutsk State University (API ISU), Gagarin Blvd. 20, Irkutsk,
664003, Russia.
}%
\author{\firstname{R.}~\surname{Mirzoyan}}
\affiliation{%
Max Planck Institute for Physics, Fohringer Ring 6, D-80805 Munich, Germany.
}%
\affiliation{%
Lomonosov Moscow State University Skobeltsyn Institute of Nuclear Physics (MSU SINP),
Leninskie gory 1(2), GSP-1, Moscow, 119991, Russia.
}%
\author{\firstname{R.~D.}~\surname{Monkhoev}}
\affiliation{%
Institute of Applied Physics, Irkutsk State University (API ISU), Gagarin Blvd. 20, Irkutsk,
664003, Russia.
}%
\author{\firstname{E.~A.}~\surname{Osipova}}
\affiliation{%
Lomonosov Moscow State University Skobeltsyn Institute of Nuclear Physics (MSU SINP),
Leninskie gory 1(2), GSP-1, Moscow, 119991, Russia.
}%
\author{\firstname{A.~L.}~\surname{Pakhorukov}}
\affiliation{%
Institute of Applied Physics, Irkutsk State University (API ISU), Gagarin Blvd. 20, Irkutsk,
664003, Russia.
}%
\author{\firstname{А.}~\surname{Pan}}
\affiliation{%
Joint Institute for Nuclear Research, Joliot-Curie 6, Dubna, Moscow Region, 141980, Russia.
}%
\author{\firstname{L.~V.}~\surname{Pankov}}
\affiliation{%
Institute of Applied Physics, Irkutsk State University (API ISU), Gagarin Blvd. 20, Irkutsk,
664003, Russia.
}%
\author{\firstname{A.~A.}~\surname{Petrukhin}}
\affiliation{%
National Research Nuclear University MEPhI (Moscow Engineering Physics Institute),
Kashirskoe highway 31, Moscow, 115409, Russia.
}%
\author{\firstname{D.~A.}~\surname{Podgrudkov}}
\affiliation{%
Lomonosov Moscow State University Skobeltsyn Institute of Nuclear Physics (MSU SINP),
Leninskie gory 1(2), GSP-1, Moscow, 119991, Russia.
}%
\author{\firstname{V.~A.}~\surname{Poleschuk}}
\affiliation{%
Institute of Applied Physics, Irkutsk State University (API ISU), Gagarin Blvd. 20, Irkutsk,
664003, Russia.
}%
\author{\firstname{E.~G.}~\surname{Popova}}
\affiliation{%
Lomonosov Moscow State University Skobeltsyn Institute of Nuclear Physics (MSU SINP),
Leninskie gory 1(2), GSP-1, Moscow, 119991, Russia.
}%
\author{\firstname{А.}~\surname{Porelli}}
\affiliation{%
Deutsches Elektronen-Synchrotron DESY, 15738 Zeuthen, Germany.
}%
\author{\firstname{E.~B.}~\surname{Postnikov}}
\affiliation{%
Lomonosov Moscow State University Skobeltsyn Institute of Nuclear Physics (MSU SINP),
Leninskie gory 1(2), GSP-1, Moscow, 119991, Russia.
}%
\author{\firstname{V.~V.}~\surname{Prosin}}
\affiliation{%
Lomonosov Moscow State University Skobeltsyn Institute of Nuclear Physics (MSU SINP),
Leninskie gory 1(2), GSP-1, Moscow, 119991, Russia.
}%
\author{\firstname{V.~S.}~\surname{Ptuskin}}
\affiliation{%
Pushkov Institute of Terrestrial Magnetism, Ionosphere and Radio Wave Propagation of the
Siberian Branch of the Russian Academy of Sciences (IZMIRAN), Kaluzhskoe highway 4,
Moscow, Troitsk, 108840, Russia.
}%
\author{\firstname{A.~A.}~\surname{Pushnin}}
\affiliation{%
Institute of Applied Physics, Irkutsk State University (API ISU), Gagarin Blvd. 20, Irkutsk,
664003, Russia.
}%
\author{\firstname{R.~I.}~\surname{Raikin}}
\affiliation{%
Altai State University, Lenina 61, Barnaul, 656049, Russia.
}%
\author{\firstname{A.}~\surname{Razumov}}
\affiliation{%
Lomonosov Moscow State University Skobeltsyn Institute of Nuclear Physics (MSU SINP),
Leninskie gory 1(2), GSP-1, Moscow, 119991, Russia.
}%
\author{\firstname{E.}~\surname{Rjabov}}
\affiliation{%
Institute of Applied Physics, Irkutsk State University (API ISU), Gagarin Blvd. 20, Irkutsk,
664003, Russia.
}%
\author{\firstname{G.~I.}~\surname{Rubtsov}}
\affiliation{%
Institute for Nuclear Research of the Russian Academy of Sciences, 60th October
Anniversary 7a, 117312, Moscow, Russia.
}%
\author{\firstname{Y.~I.}~\surname{Sagan}}
\affiliation{%
Joint Institute for Nuclear Research, Joliot-Curie 6, Dubna, Moscow Region, 141980, Russia.
}%
\affiliation{%
Dubna State University, Universitetskaya 19, Dubna, Moscow region, 141982, Russia.
}%
\author{\firstname{V.~S.}~\surname{Samoliga}}
\affiliation{%
Institute of Applied Physics, Irkutsk State University (API ISU), Gagarin Blvd. 20, Irkutsk,
664003, Russia.
}%
\author{\firstname{A.~Yu.}~\surname{Sidorenkov}}
\affiliation{%
Institute for Nuclear Research of the Russian Academy of Sciences, 60th October
Anniversary 7a, 117312, Moscow, Russia.
}%
\author{\firstname{A.~А.}~\surname{Silaev}}
\affiliation{%
Lomonosov Moscow State University Skobeltsyn Institute of Nuclear Physics (MSU SINP),
Leninskie gory 1(2), GSP-1, Moscow, 119991, Russia.
}%
\author{\firstname{A. A.}~\surname{Silaev (junior)}}
\affiliation{%
Lomonosov Moscow State University Skobeltsyn Institute of Nuclear Physics (MSU SINP),
Leninskie gory 1(2), GSP-1, Moscow, 119991, Russia.
}%
\author{\firstname{A.~V.}~\surname{Skurikhin}}
\affiliation{%
Lomonosov Moscow State University Skobeltsyn Institute of Nuclear Physics (MSU SINP),
Leninskie gory 1(2), GSP-1, Moscow, 119991, Russia.
}%
\author{\firstname{I.}~\surname{Satyshev}}
\affiliation{%
Joint Institute for Nuclear Research, Joliot-Curie 6, Dubna, Moscow Region, 141980, Russia.
}%
\author{\firstname{A.~V.}~\surname{Sokolov}}
\affiliation{%
Novosibirsk State University, Pirogova 1, Novosibirsk, 630090, Russia.
}%
\affiliation{%
Budker Institute of Nuclear Physics of the Siberian Branch of the Russian Academy of
Sciences, Lavrentyev Prosp. 11, Novosibirsk, 630090, Russia.
}%
\author{\firstname{Y.}~\surname{Suvorkin}}
\affiliation{%
Institute of Applied Physics, Irkutsk State University (API ISU), Gagarin Blvd. 20, Irkutsk,
664003, Russia.
}%
\author{\firstname{L.~G.}~\surname{Sveshnikova}}
\affiliation{%
Lomonosov Moscow State University Skobeltsyn Institute of Nuclear Physics (MSU SINP),
Leninskie gory 1(2), GSP-1, Moscow, 119991, Russia.
}%
\author{\firstname{V.~A.}~\surname{Tabolenko}}
\affiliation{%
Institute of Applied Physics, Irkutsk State University (API ISU), Gagarin Blvd. 20, Irkutsk,
664003, Russia.
}%
\author{\firstname{A.~B.}~\surname{Tanaev}}
\affiliation{%
Institute of Applied Physics, Irkutsk State University (API ISU), Gagarin Blvd. 20, Irkutsk,
664003, Russia.
}%
\author{\firstname{В.~А.}~\surname{Tarashansky}}
\affiliation{%
Institute of Applied Physics, Irkutsk State University (API ISU), Gagarin Blvd. 20, Irkutsk,
664003, Russia.
}%
\author{\firstname{M.}~\surname{Ternovoy}}
\affiliation{%
Institute of Applied Physics, Irkutsk State University (API ISU), Gagarin Blvd. 20, Irkutsk,
664003, Russia.
}%
\author{\firstname{L.~G.}~\surname{Tkachev}}
\affiliation{%
Joint Institute for Nuclear Research, Joliot-Curie 6, Dubna, Moscow Region, 141980, Russia.
}%
\affiliation{%
Dubna State University, Universitetskaya 19, Dubna, Moscow region, 141982, Russia.
}%
\author{\firstname{M.}~\surname{Tluczykont}}
\affiliation{%
Institute of experimental physics of Hamburg University, Luruper Chaussee 149, 22761
Hamburg, Germany.
}%
\author{\firstname{N.}~\surname{Ushakov}}
\affiliation{%
Institute for Nuclear Research of the Russian Academy of Sciences, 60th October
Anniversary 7a, 117312, Moscow, Russia.
}%
\author{\firstname{А.}~\surname{Vaidyanathan}}
\affiliation{%
Novosibirsk State University, Pirogova 1, Novosibirsk, 630090, Russia.
}%
\author{\firstname{P.~A.}~\surname{Volchugov}}
\affiliation{%
Lomonosov Moscow State University Skobeltsyn Institute of Nuclear Physics (MSU SINP),
Leninskie gory 1(2), GSP-1, Moscow, 119991, Russia.
}%
\author{\firstname{N.~V.}~\surname{Volkov}}
\affiliation{%
Altai State University, Lenina 61, Barnaul, 656049, Russia.
}%
\author{\firstname{D.}~\surname{Voronin}}
\affiliation{%
Institute for Nuclear Research of the Russian Academy of Sciences, 60th October
Anniversary 7a, 117312, Moscow, Russia.
}%
\author{\firstname{R.}~\surname{Wischnewski}}
\affiliation{%
Deutsches Elektronen-Synchrotron DESY, 15738 Zeuthen, Germany.
}%
\author{\firstname{I.~I.}~\surname{Yashin}}
\affiliation{%
National Research Nuclear University MEPhI (Moscow Engineering Physics Institute),
Kashirskoe highway 31, Moscow, 115409, Russia.
}%
\author{\firstname{A.~V.}~\surname{Zagorodnikov}}
\affiliation{%
Institute of Applied Physics, Irkutsk State University (API ISU), Gagarin Blvd. 20, Irkutsk,
664003, Russia.
}%
\author{\firstname{D.~P.}~\surname{Zhurov}}
\affiliation{%
Institute of Applied Physics, Irkutsk State University (API ISU), Gagarin Blvd. 20, Irkutsk,
664003, Russia.
}%


\begin{abstract}
A wide-angle Cerenkov array TAIGA-HiSCORE (FOV $\sim$0.6 sr), was originally created as a part of TAIGA installation for high-energy gamma-ray astronomy and cosmic ray physics. Array now consist on nearly 100 optical stations on the area of 1 km$^2$. Due to high accuracy and stability ($\sim$1 ns) of time synchronization of the optical stations the accuracy of EAS arrival direction reconstruction is reached 0.1$^\mathrm{o}$. It was proven that the array can also be used to search for nanosecond events of the optical range. The report discusses the method of searching for optical transients using the HiSCORE array and demonstrates its performance on a real example of detecting signals from an artificial Earth satellite. The search for this short flares in the HiSCORE data of the winter season 2018--2019 is carried out. One candidate for double repeater has been detected, but the estimated probability of random simulation of such a transient by background EAS events is not less than 10\%, which does not allow us to say that the detected candidate corresponds to a real astrophysical transient. An upper bound on the frequency of optical spikes with flux density of more than $10^{-4}\,\mathrm{erg/s/cm}^2$ and a duration of more than 5\,ns is established as $\sim 2 \times 10^{-3}$\,events/sr/hour.
\end{abstract}

\maketitle

\section{Introduction}

The astrophysical complex TAIGA (Tunka Advanced Instrument for cosmic ray physics and Gamma-ray Astronomy) \cite{TAIGA-NIM-2017A,TAIGA-NIM-2017B,TAIGA-NIM-2020A,TAIGA-NIM-2020B} is located in the Tunka valley at 50 km from Baikal Lake and is designed for the study of cosmic rays physics and gamma-ray astronomy of high and ultrahigh energies using a technique based on the observation of extensive air showers (EAS). The complex includes several instruments, including the TAIGA-HiSCORE wide-aperture Cherenkov array. The operating principle of the TAIGA-HiSCORE telescope is based on the idea presented in the article \cite{HISCORE-2011}: the detector stations measure the amplitudes of the light signal caused by the Cherenkov light from a EAS, as well as the time evolution of the Cherenkov light front reconstructed according to the sequence of stations triggering.  The HiSCORE array (High Sensitivity COsmic Rays and gamma Explorer) consists of integrated optical atmospheric Cherenkov stations, each of which includes four PMTs with a total entrance pupil area of about 0.5\,m${}^2$ and a FOV of 0.6~steradian, pointed to the same sky direction, and form an array with distance between two stations is 106 m (at present they are adapted for observing the Crab Nebula). The HiSCORE array is in the process of construction, and currently consists of 120 stations located on area of more than 1\,km${}^2$. New Cherenkov stations, as they were added to the telescope, were immediately put into operation, so the instrument was taking data during its construction. The design and construction details of the HiSCORE telescope are presented in the articles \cite{HISCORE-2011,TAIGA-NIM-2017B}. It is possible to reconstruct the direction of arrival of a cosmic ray particle (charged or gamma quantum) from this data, its initial energy, and some other important characteristics of EAS.

\begin{figure}
 \begin{center}
 \includegraphics[width=\htw]{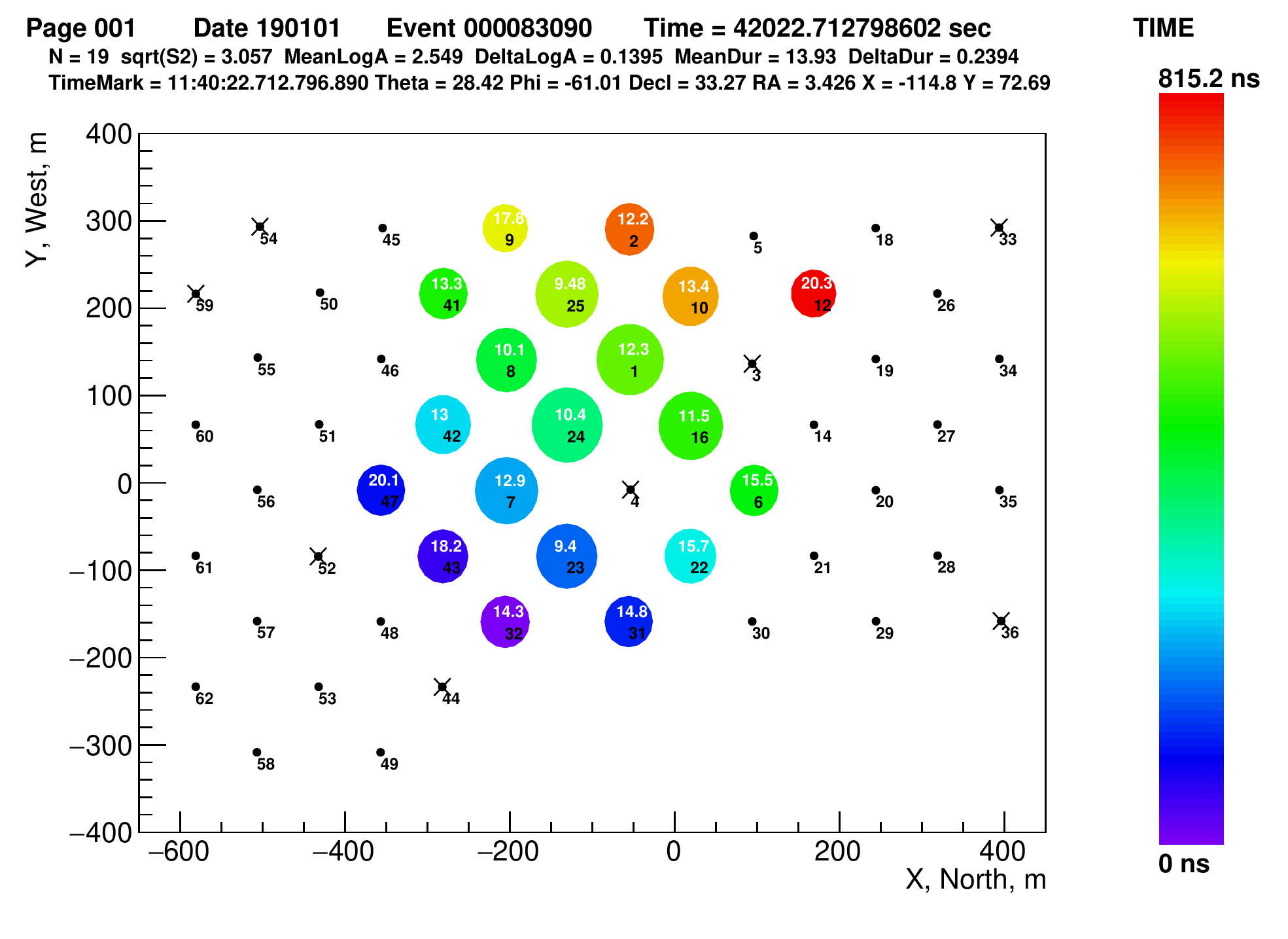}
 \includegraphics[width=\htw]{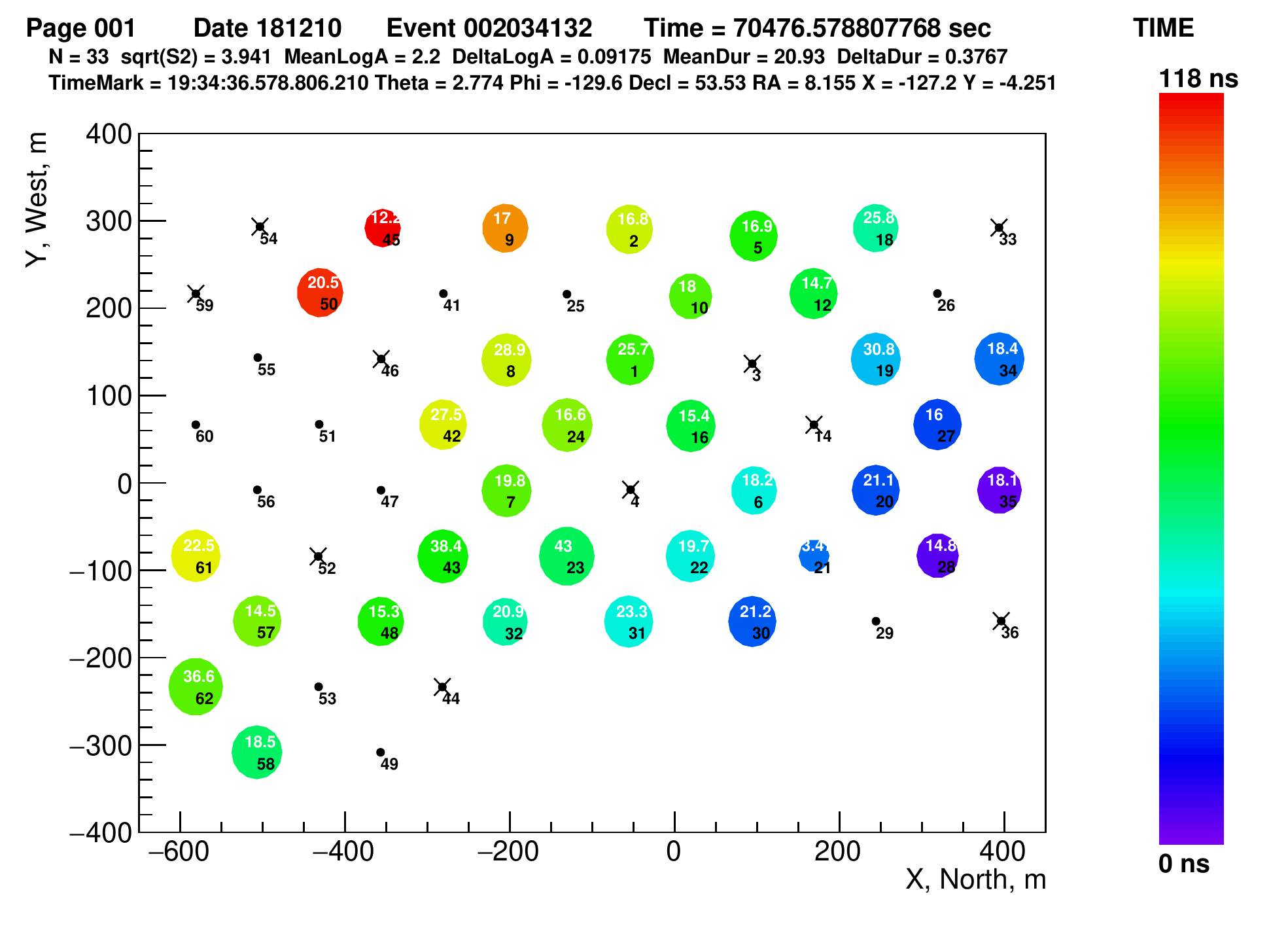}
 \end{center}
\caption{\label{fig:Portrait} 
Portraits of a typical EAS event (left diagram) and an event of a distant optical transient, which is represented by a laser pulse (LIDAR) from a distance of about 700\,km. The diagrams show the response times of the optical stations in color, and the estimates of the signal duration in nanoseconds are inscribed in the colored circles of the triggered stations in numbers. The size of the circles represents the amplitude of the signal in a logarithmic scale. The title of each portrait contains a variety of technical information about the event. The numbers of the stations of the HiSCORE telescope are also shown, while the crosses indicate the stations that at the time of measuring of the event for some reason were not functioning or had operational problems. The configuration of the HiSCORE array corresponds to the winter measurement season of 2018--2019.}
\end{figure}

\begin{figure}
 \begin{center}
 \includegraphics[width=\tw]{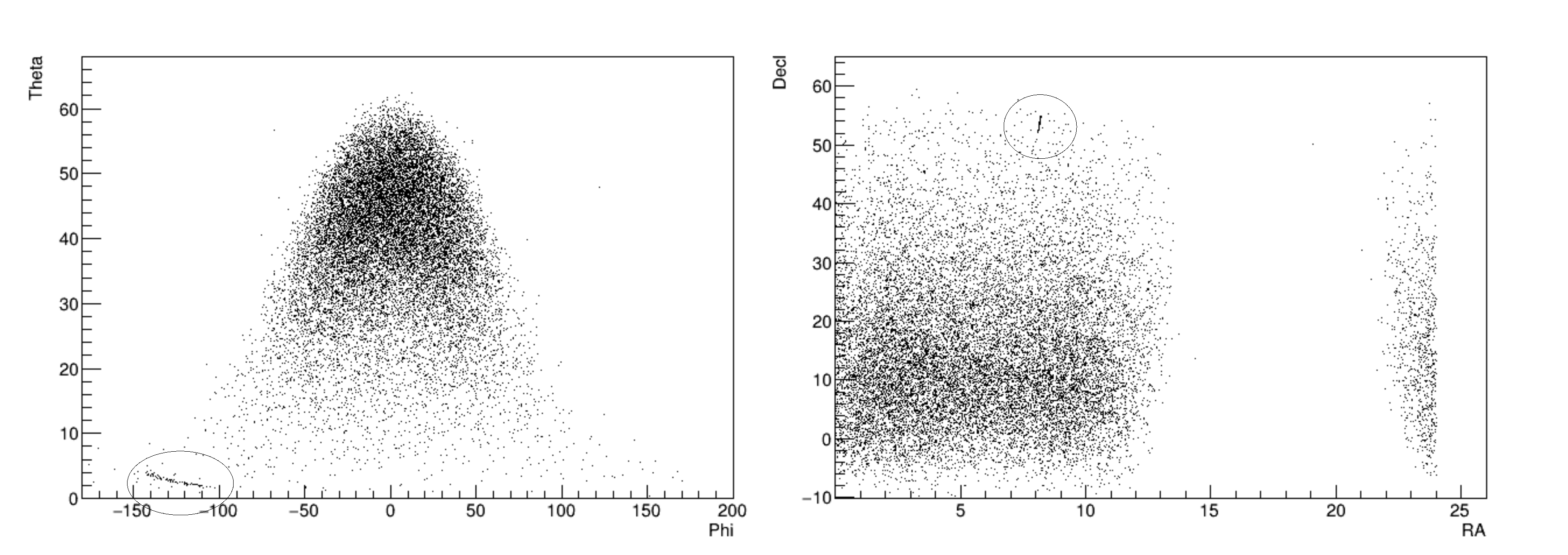}
 \end{center}
\caption{\label{fig:CALIPSO} 
Local and equatorial coordinates of all events with with number of triggered stations $N_{\mathrm{Eff}} \ge 20$, detected on the night of December 10, 2018 universal time. Outlined points are the track of satellite events.}
\end{figure}

The characteristic duration of EAS Cherenkov light pulses ranges from several nanoseconds to tens of nanoseconds; therefore, the HiSCORE detector stations are adapted to register just such flashes. The detection threshold is approximately 3000 quanta of green region ($\sim$530\,nm) per square meter during an integration time of 10\,ns. However, in principle, the array is capable of registering not only EAS Cherenkov light pulses, but any flashes of light. In this case, according to a number of characteristics, a flash of a distant point source (hereinafter referred to as \emph{optical transient\/}) in the great majority of cases can be easily distinguished from a passage of a Cherenkov front of a EAS. Indeed, the stations triggered by an EAS Cherenkov light pulse will distribute around the shower axis, so that the maximum signals from triggered stations will be closest to the axis and the event will look like a relatively compact "spot" near the EAS axis. On the contrary, distant transient events generate a flat and homogeneous optical front that evenly covers the entire HiSCORE array, so that all optical stations receive the same optical pulse at the input within the statistical accuracy. Therefore, it should be expected that the event of a distant transient will illuminate the HiSCORE array, uniformly distributed over the installation area, and there will be no pronounced dependence of the amplitude of the triggered station on its position. This idea is confirmed by comparing the structure of a typical EAS event and an event caused by a pulse of a laser installed on the orbital satellite (\fig{Portrait}).
The signal from the satellite flying over the HiSCORE array was unexpectedly detected in the data for December 10, 2018 (see \fig{CALIPSO}) and was subsequently identified as the signal from the LIDAR on the CALIPSO satellite \cite{TAIGA-ICRC-2021A}. Further, it was found that the signal from CALIPSO has been present in the data since 2015 and can be used for the absolute pointing calibration of HiSCORE array \cite{TAIGA-ICRC-2021A} just as it was obtained using signals from the lidar on ISS \cite{TAIGA-ICRC-ISS-2017}. The source of the optical pulse on the satellite is distant enough to generate an almost flat and uniform optical front near HiSCORE, and well simulates an infinitely distant source. \fig{Portrait} clearly shows the difference between the structure of EAS events and the satellite events. The optical transient events easily pass the primary selection of events in the main working method of HiSCORE, therefore, the program for searching for transients in the HiSCORE data can be implemented in a pure concomitant mode. The task is reduced only to the implementation of some specific HiSCORE event filtering algorithms. Thus, the HiSCORE Cherenkov array, originally adapted for the study of cosmic ray physics and for gamma-ray astronomy, can be relatively easily adopted for solving problems of conventional optical astronomy.

An interesting question is what kind of distant astrophysical sources can lead to nanosecond flashes of light that could be detected by the HiSCORE apparatus. If the source is incoherent, then the size of the emitter should be measured at most in units of meters, since in 1\,ns light passes only 0.3\,m. In principle, such compact incoherent emitters can be the remnants of evaporating relic black holes. If the source of the pulse is coherent, then there are no restrictions on the size of the source, but in this case we should talk about a quantum generator of either natural or artificial origin. Natural cosmic sources of coherent radiation, including optical lasers, are known \cite{LETOKHOV-2002,LETOKHOV-2007}, and they emit in a continuous mode. Natural sources of short coherent optical pulses is unknown and require further studies. The most understandable hypothetical source of short optical pulses is a distant optical laser of artificial origin. 

The use of lasers for interstellar communication was proposed by R. N. Schwartz and C. H. Townes in 1961 \cite{SCHWARTZ-1961}. Thus, the task is within the framework of the SETI (Search for Extra Terrestrial Intelligence). It is easy to understand that even for very large distances, a laser with relatively modest parameters is sufficient for its radiation to be detected by HiSCORE optical stations. For example, for a transmitter with an aperture of 1000\,m, for light with a wavelength of 530\,nm (green), at a distance of $10^4$\,ly from the Solar System, the diffraction radius of the light spot will be about 50\,million\,km, from where for a threshold of 3000 quanta/m$^2$/10\,ns we obtain that the laser energy should be about 9\,MJ for every 10\,ns. Such energy and power are comparable to the lasers already used on Earth in thermonuclear fusion programs. An aperture of 1000\,m means the use of a laser phased array. Prototypes of such devices also already exist on Earth. These estimates show that, using very modest laser transmitters and modest optical receivers, the transmission of information through an optical channel can, in principle, be established at virtually any intragalactic distances.

Note that the idea of using large Cherenkov telescopes to search for optical transients of astrophysical origin was proposed already at the very beginning of the 2000s \cite{CH-BESKIN-2001}, including also the framework of the SETI problem \cite{CH-SETI-EICHLER-BESKIN-2001}. After that, the idea was repeatedly discussed in various aspects and some search programs were implemented \cite{CH-SETI-HOLDER-2005,CH-GRIFFIN-2011,CH-BARTOS-2014,CH-BARTOS-2018,CH-MAGIC-2018}. However, in all these works, it was assumed and discussed the use of image Cherenkov telescopes, which have a small field of view, measured by angles of only a few degrees. In the present work, for the first time, an instrument of a completely different type is used for the purposes of optical astronomy. The HiSCORE Cherenkov array has a very wide field of view, which covers about one tenth of the visible sky. This is extremely important in the SETI when it comes to detecting very rare events. 

\section{Methods}

In this work, we study the data of the HiSCORE array recorded in the winter season from November 2018 to April 2019, when the TAIGA-HiSCORE array included 54 optical stations (according to astroclimatic conditions, only winter seasons are the working time of the HiSCORE array). The time of operation consisted of 80 clear nights with a total observation time of 475 hours and an exposure of 288~sr$\cdot$hour.

\begin{figure}
 \begin{center}
 \includegraphics[width=\htw]{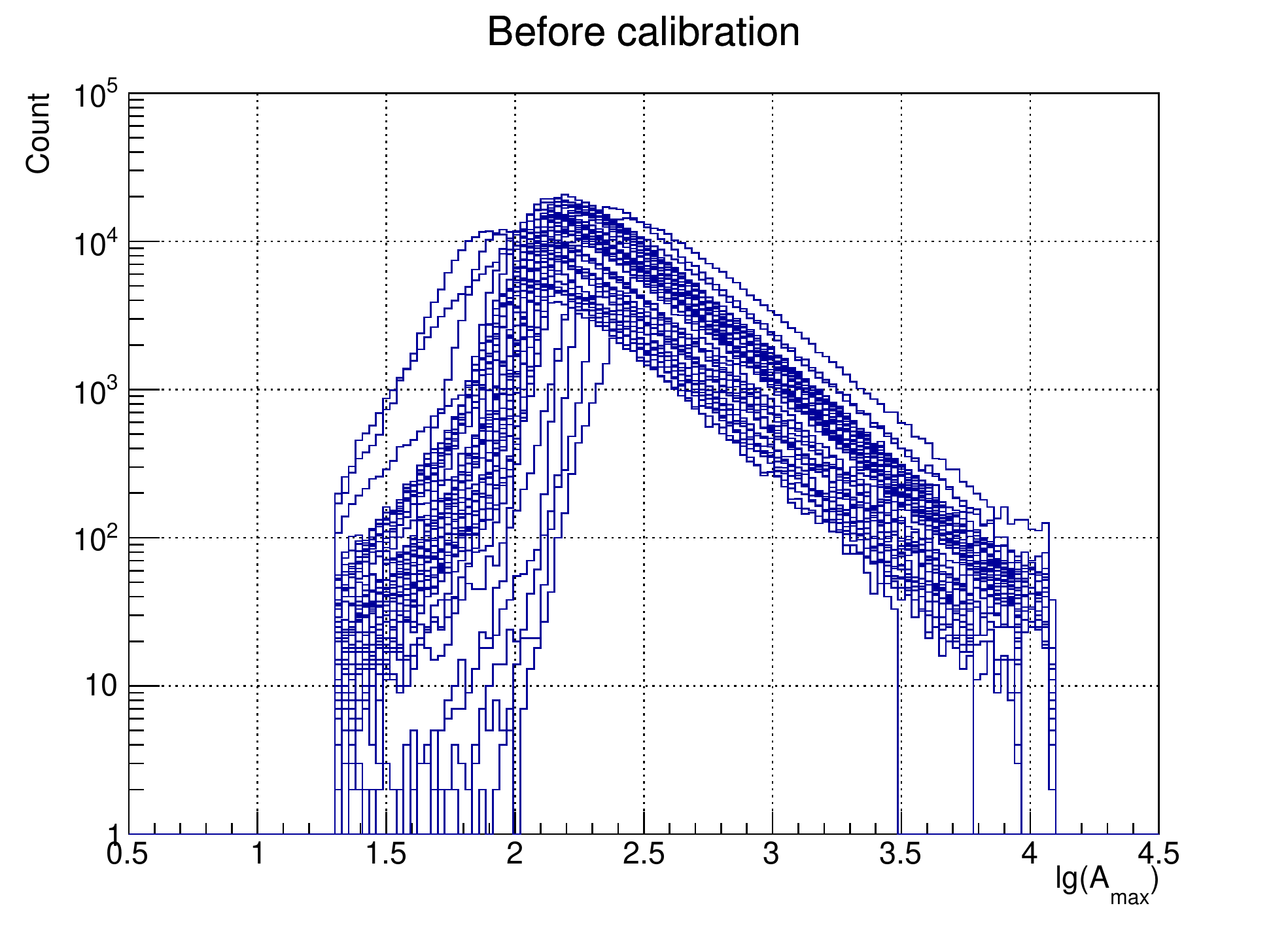}
 \includegraphics[width=\htw]{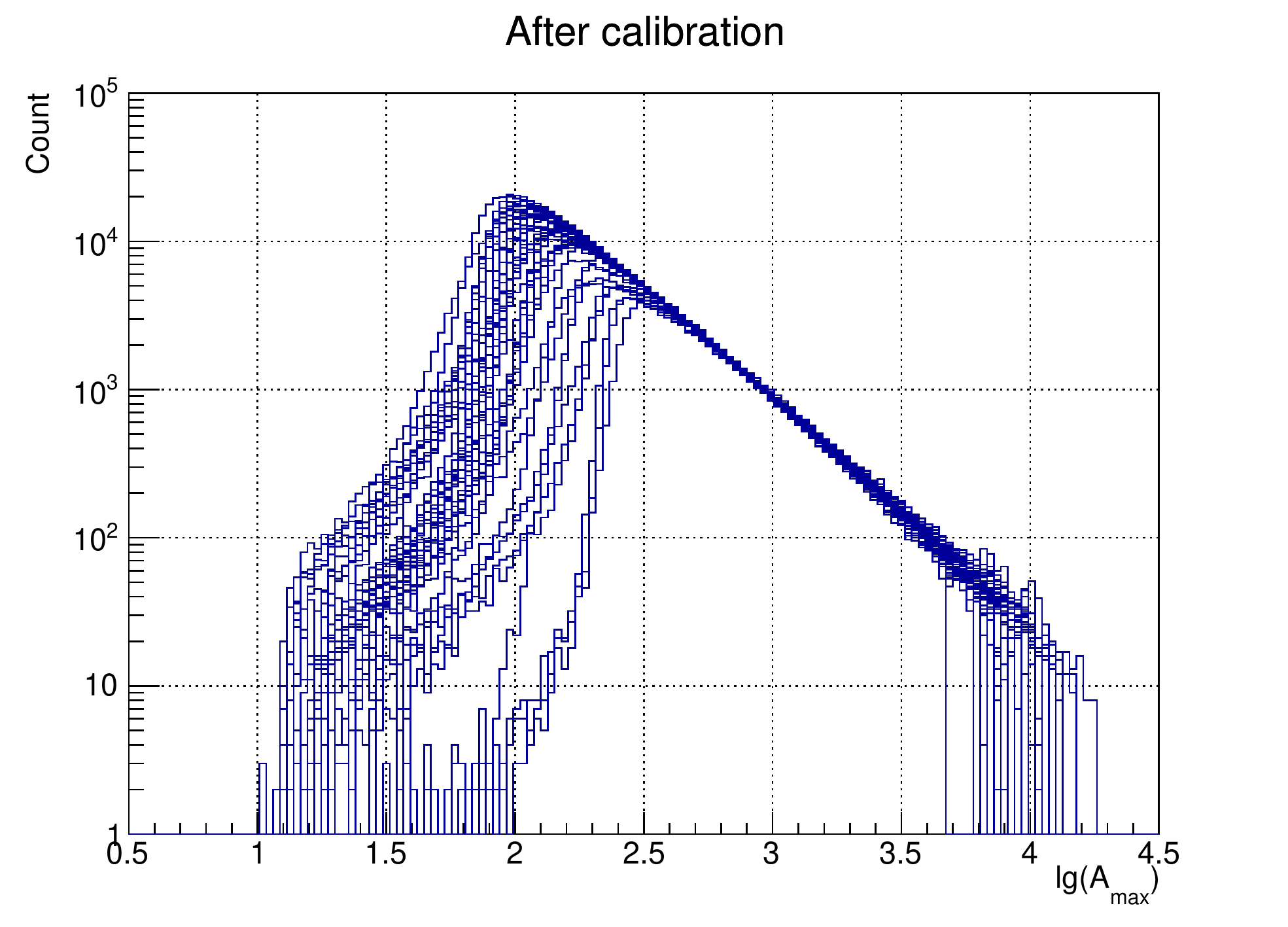}
 \end{center}
\caption{\label{fig:Calibration} Spectra of maximum amplitudes before the calibration of the optical stations amplification and after calibration.}
\end{figure}

If a triggering of optical stations is caused by a distant optical transient, then, with the same sensitivity of optical stations, the maximum pulse amplitudes of all stations should ideally be the same up to statistical fluctuations. For this reason, one of the important criteria in the search for candidates for optical transients is the amplitude comparison between stations. However, the amplification factors of different stations are not quite the same; therefore, a special relative amplitude calibration is required, which makes it possible to compare the amplitudes of different stations of the same event with each other. The idea behind this calibration is very simple. During each night of measurements, all optical stations are in the same conditions, having on average the same flux of EAS optical pulses at the input. Therefore, equally operating stations should show the same amplitude spectra. In reality, this is not the case due to differences in the amplification factors of the stations. The idea of calibration is to select such a correction factor for the amplitudes for each optical station so that all spectra of the stations are reduced to some single average spectrum, the same for all stations. After this is done, each station gives the same response to the same input fluxes of optical pulses, therefore, the stations operate effectively in the same way and the amplitudes of the pulses of different stations can be adequately compared with each other. \fig{Calibration} shows the amplitude spectra of all optical stations for one observation night before and after calibration (calibration is performed for each observation night separately). It can be seen that after calibration, the amplitude spectra of the stations achieve to the matching level, except that different stations have different thresholds. This effect, unfortunately, is unavoidable. In the case of the arrival of a flat optical front of a distant transient, generally speaking, not all stations will be triggered, since for some stations the optical pulse may turn out to be higher than their threshold, while for some other stations -- lower. However, the event will still demonstrate a small spread in the amplitudes of different triggered stations and the absence of a concentration of triggered stations around any axis, as is the case in EAS events.

The most important task of HiSCORE data processing is the reconstruction of the direction to the source, which is carried out using the measured response times of different optical stations caused by the passage of the light front. In this work, the direction is reconstructed under the assumption of a plane light front, which is absolutely correct for a distant optical transient, but also gives the direction of the EAS axis, despite the fact that the shape of the front of EAS Cherenkov light is not completely flat. This is possible due to the fact that, although the EAS front is not flat, it is axially symmetric with respect to the shower axis. The problem of finding the direction to the source is reduced to minimization along two angular coordinates -- $\varphi$ (azimuth) and $\theta$ (polar angle) -- of the function
\begin{equation}
 W(\theta,\varphi) = \sum_{i=1}^N
 \left[
 T(\theta,\varphi) + \frac{1}{c}(x_i\sin\theta\cos\varphi + 
 y_i\sin\theta\sin\varphi + z_i\cos\theta) - t^*_i
 \right]^2, 
 \label{eq:W}
\end{equation}
where $N$ is the number of triggered optical stations, $x_i,y_i,z_i$ are the coordinates of the triggered stations, $t^*_i$ are the trigger times of the stations, $c$ is the speed of light and
\begin{equation}
 T(\theta,\varphi) = 
 \frac{1}{N}\sum_{i=1}^N
 \left[
 t_i^* - 
 \frac{1}{c}(x_i\sin\theta\cos\varphi + 
 y_i\sin\theta\sin\varphi + z_i\cos\theta)
 \right].
 \label{eq:T-theta-phi}
\end{equation}

Several passes through the HiSCORE aperture of the CALIPSO satellite have been recorded \cite{TAIGA-ICRC-2021A}. The satellite's orbital altitude is 700\,km, the laser pulse duration is about 20\,ns, therefore CALIPSO provides an almost ideal test for signals from a distant optical transient. The scatter of the trajectory points was used to determine the accuracy of the reconstruction of the angular coordinates for events near the zenith (where the satellite appears in the HiSCORE aperture), which was about 0.05$^{\mathrm{o}}$ \cite{TAIGA-ICRC-2021A}. As mentioned in the Introduction, the observation of LIDAR events also confirmed the idea of filtering events of distant optical transients from the EAS background (see \fig{Portrait}).

\section{Results}

\begin{figure}
\begin{center}
\includegraphics[width=\htw]{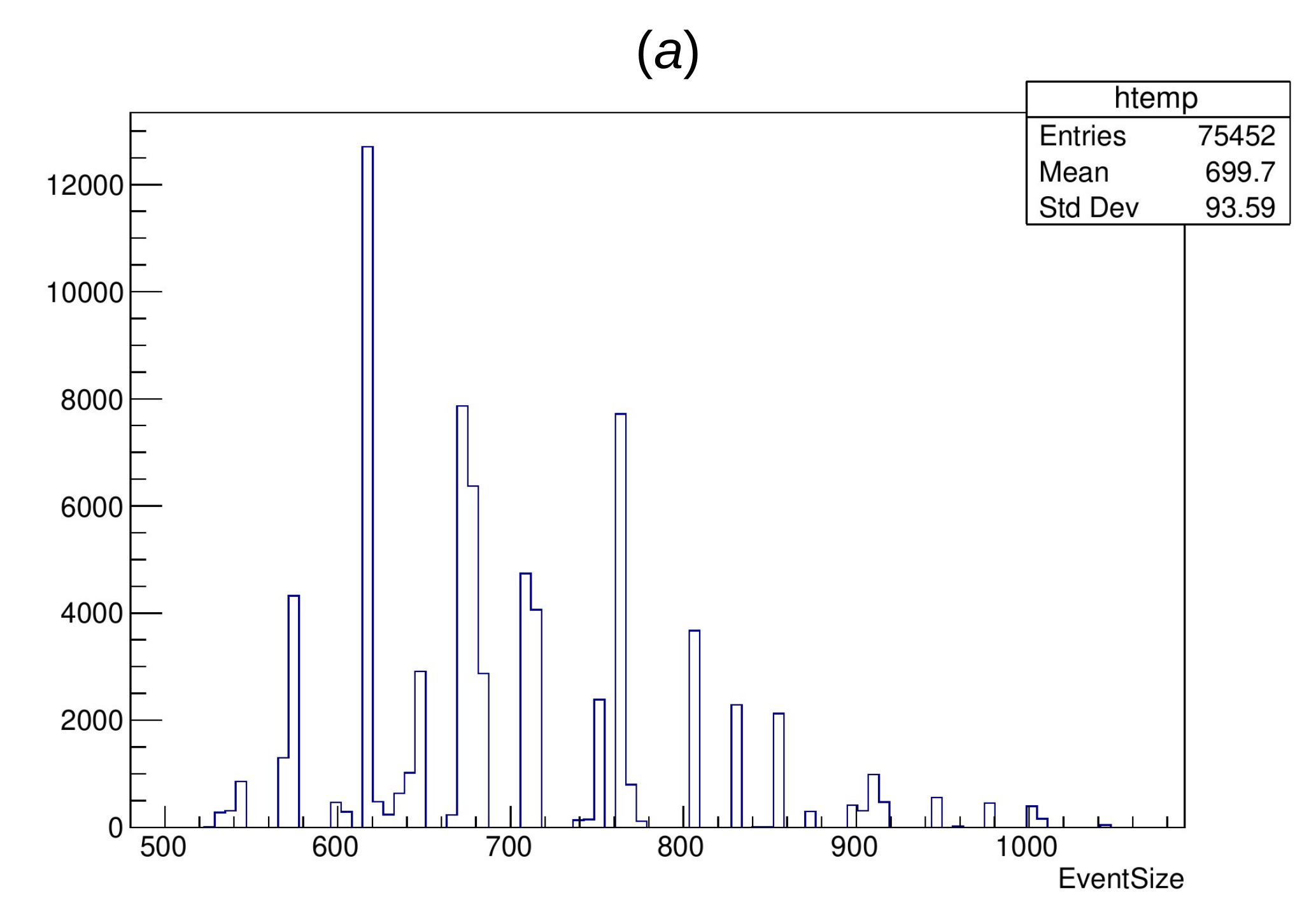}
\includegraphics[width=\htw]{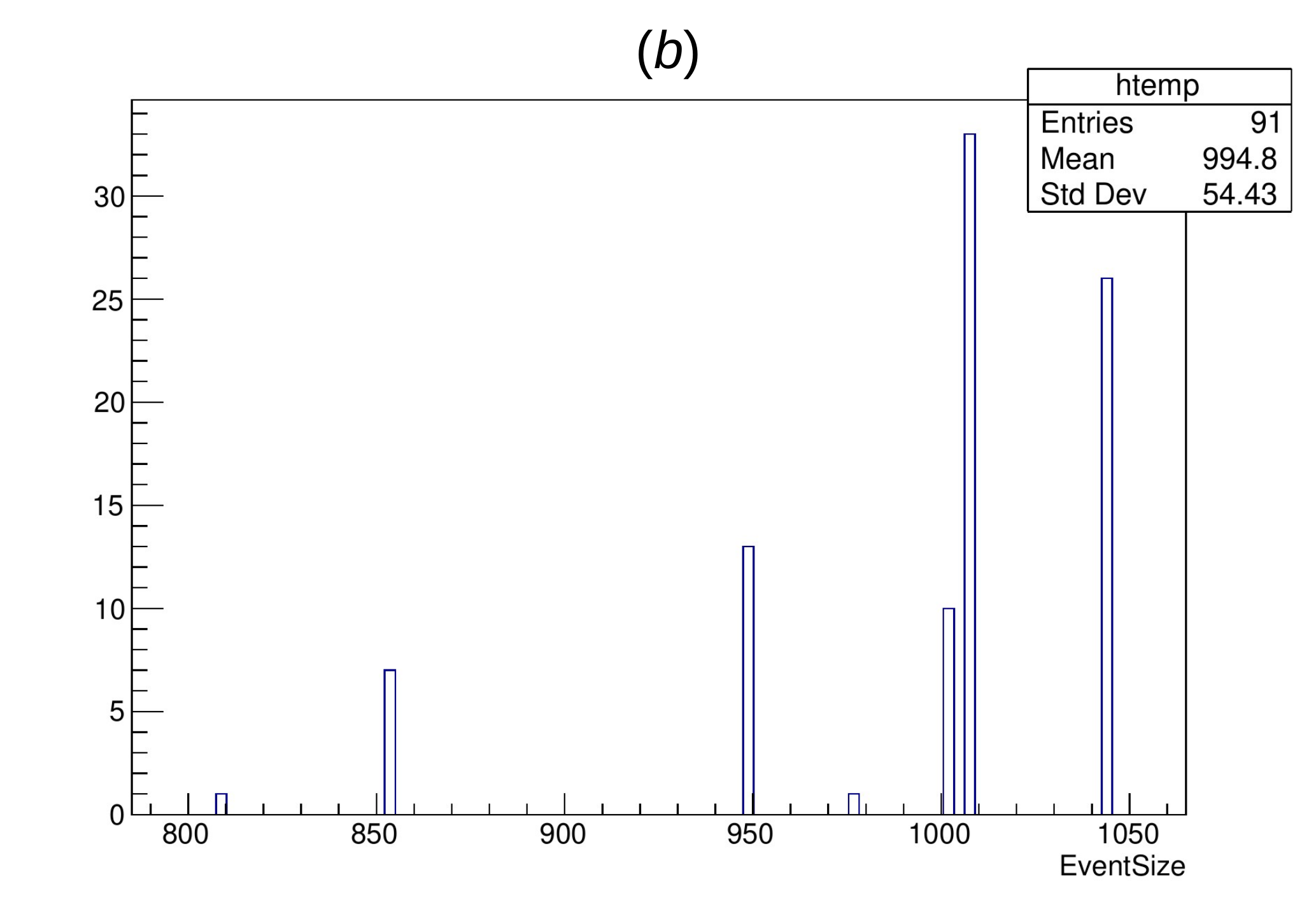}
\end{center}
\caption{Distribution of events by parameter ``event size'' (\texttt{EventSize}, m) for all events with trigger multiplicity $N_{\mathrm{Eff}} \ge 20$ and the condition \texttt{DeltaLogA < 0.1} without satellite events ($a$) and for satellite events only ($b$).
\label{fig:EventSize-Prob}}
\end{figure}

Since the event of a distant optical transient covers the entire area of the HiSCORE array uniformly, in contrast to the great majority of EAS events, it is expected that a certain parameter representing the degree of uniformity of illumination of the array area can be effectively used to select candidates for distant transients. As such a parameter, we introduce a parameter called \texttt{EventSize}. The event size is understood as the maximum distance in the $(x,y)$ plane in meters between the triggered stations belonging to this event. If this criterion is used in conjunction with the condition \texttt{DeltaLogA < 0.1}, where \texttt{DeltaLogA} is the standard deviation for the logarithm of the amplitude of triggered stations for one event, then it turns out to be very effective for selecting potential candidates for distant transients. The limitation on the spread of amplitudes for optical transient events \texttt{DeltaLogA < 0.1} was deduced from the satellite test events (see the Methods section above). The left panel of \fig{EventSize-Prob} shows the distribution by the \texttt{EventSize} parameter for all events with the number of triggered stations $N_{\mathrm{Eff}} \ge 20$ and with the additional condition \texttt{DeltaLogA < 0.1}, except for satellite events; the right panel shows the same distribution for satellite events only. It can be seen that the great majority of satellite events are larger than 900~m, while the great majority of all other events are less than 900~m with a maximum probability near 600~m. Therefore, if we consider events with a size greater than 900~m, then we will get rid of most of the EAS background, preserving most of the satellite events (i.e., most of the events of distant transients).

\begin{figure}
\begin{center}
\includegraphics[width=\htw]{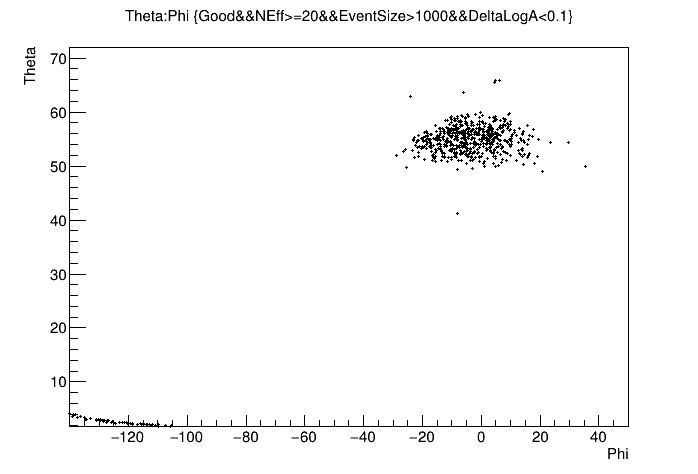}
\includegraphics[width=\htw]{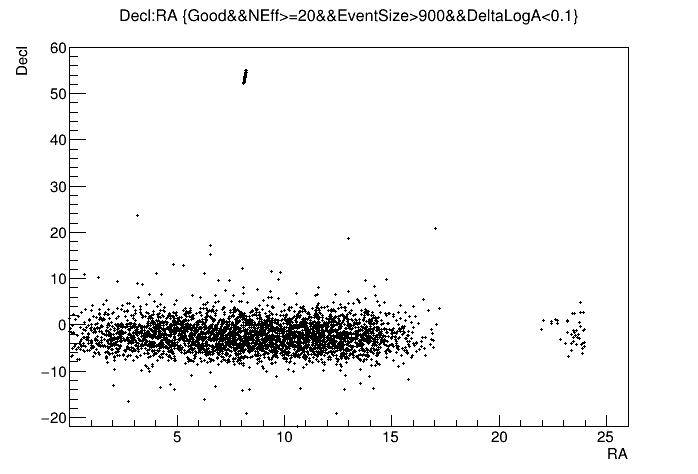}
\end{center}
\caption{Distribution of events over the sky for the events with a size of more than 900~m and the condition \texttt{DeltaLogA < 0.1}: in the region of small zenith angles (or large declination angles) there are no events other than satellite events.
\label{fig:EventSize-Coord}}
\end{figure}

\fig{EventSize-Coord} shows the distribution of all events $N_{\mathrm{Eff}} \ge 20$, \texttt{EventSize > 900~m}, \texttt{DeltaLogA < 0.1} over the sky in local and equatorial coordinates. In the region of small zenith angles, $\theta < 40^\mathrm {o}$ (or large declination angles, \texttt{Decl > 25}$^\mathrm {o}$), a track of satellite events is visible, the great majority of which pass this filtering, being distant optical transients, but no other events. This means that there are no other candidates for distant optical transients, except for satellite events, in the region of these angles. This is one of the results of this work. The second conclusion from \fig{EventSize-Coord} is that the background of EAS events in this region of angles for searching for candidates for distant transients is very small, which is important for further studies. Virtually any event detected here should be considered to be a distant optical transient candidate.

However, it can be seen from the same \fig{EventSize-Coord} that in the area of large zenith angles, which means the area of low declinations in the absolute equatorial coordinates (including negative declinations), there is a large group of events that pass all filtration conditions for ``flatness''. These are rare very gentle showers that can cover the entire area of the HiSCORE array, simultaneously creating close amplitudes of different triggered optical stations, completely simulating the passage of a uniform flat front. A total of 511 such events were filtered. It is obvious that these EAS events are the background for potential events of distant optical transients. The question is, \emph{whether is it possible to find candidate events for optical transients in such a background, separating them in some way from the background events?}

The answer to this question is ``yes'' if one looks not for single candidates for transients, but for repeater candidates, where a repeater is understood as a group of events that came from one point in the sky. There is no reason other than pure chance for ordinary EAS events to come from one point, so repeated events with the same coordinates in the sky must mean the presence of a real repeating source. Finding a repeater among the filtered 511 candidates would mean finding a real candidate for a repeating optical transient if it is shown that the random simulation of such a repeater by the EAS background is small.

Obviously, by events that came from one point in the sky, one should understand such events for which the difference in angular coordinates is less than the errors in measuring of the coordinates. As mentioned above, an error in determining the angular coordinates of 0.05$^{\mathrm{o}}$ was measured for events caused by a laser mounted on a satellite. However, this result refers to events close to the zenith of the local coordinate system, while all 511 candidates are events with large zenith angles. Such events have a completely different space-time structure, and the error in determining the coordinates for them will, generally, be different. Since we do not have a natural standard for flat events with large zenith angles, the coordinate error for such events was calculated by the Monte Carlo method according to the known distribution laws of errors in the trugger times of optical stations. As a result, the value $\Delta\varphi = 0.03^{\mathrm{o}}$ was obtained, which has the meaning of the standard deviation in any of the angular coordinates. It can be seen that the error obtained is indeed different from the error for events near the zenith. An additional problem is that the distribution law of coordinate deviations is quite different from the Gaussian one, which makes it difficult to determine the corridor for indistinguishable coordinates of events. For the final calculations, we postulated that a group of events with the same coordinates are those for which all events are included in a circle with a center coinciding with the center of mass of all events in the group and a radius of $R = \Delta\varphi\sqrt{2} = 0.041^{\mathrm {o}}$.

\begin{figure}
\begin{center}
\includegraphics[width=\tw]{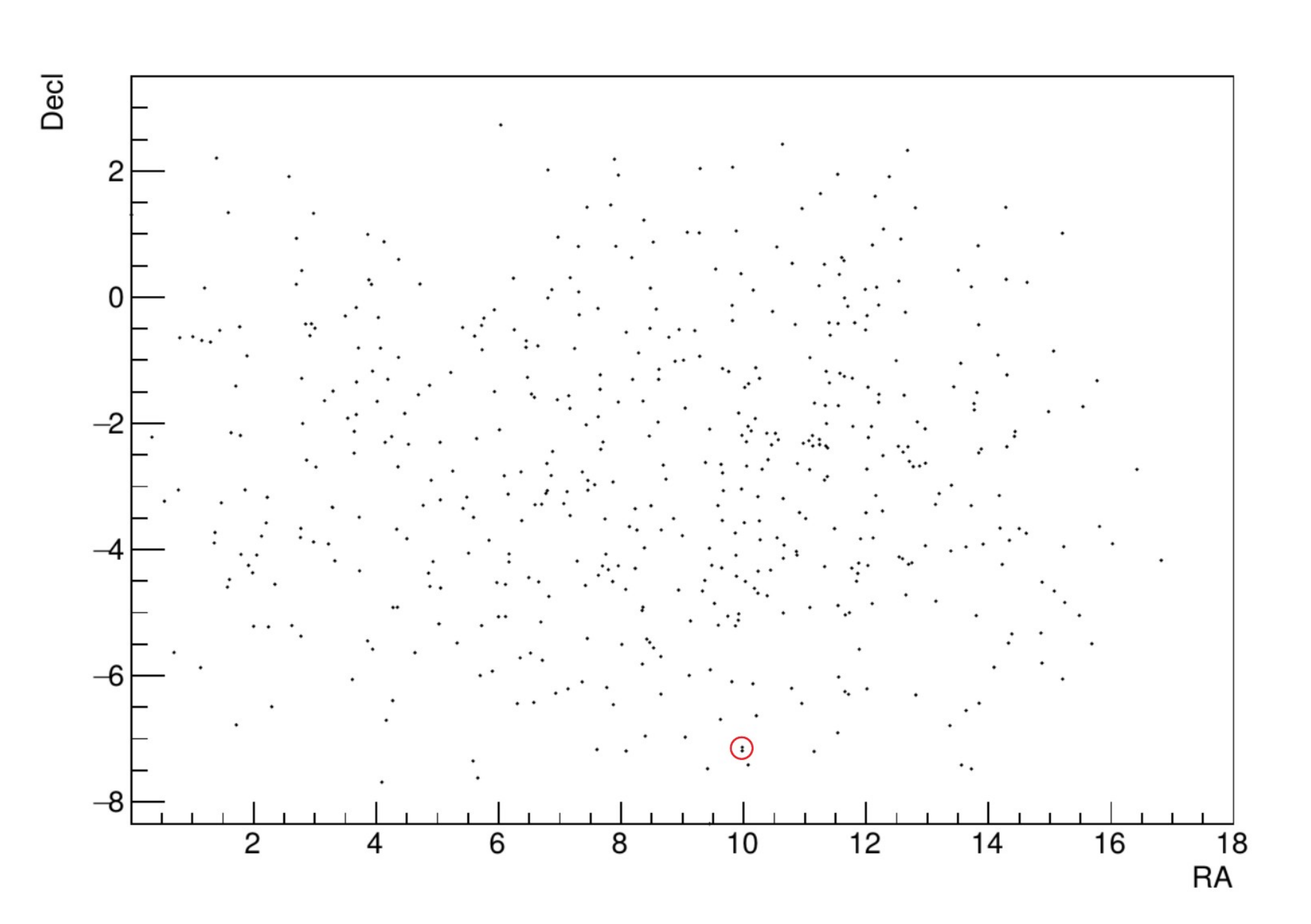}
\end{center}
\caption{Sky distribution of 511 events -- candidates for the search for optical transient repeaters. The red circle indicates one candidate found -- a double repeater.
\label{fig:Repeater}}
\end{figure}

\fig{Repeater} shows the sky distribution of all 511 events -- candidates for the search for optical transients-repeaters. The search resulted in the discovery of one pair of events with formally indistinguishable coordinates, which are marked with a red circle in the figure. Calculations by the Monte Carlo method showed that the probability of chance simulation of at least one double repeater for the distribution of events like in \fig{Repeater}, is 78\%. That is, the appearance of one double repeater should be expected simply by chance, therefore we cannot consider the found repeater as a real candidate for optical transients.

Using Monte Carlo simulations, it is possible to understand what kind of events would lead to reliable detection of a repeater candidate under the same conditions. For example, the probability of chance simulating of one triple repeater with EAS bacground is 0.002, that is, the reliability of such a candidate would be about three sigma. The likelihood of simulating a quadruple repeater is low, which would mean a virtually reliable transient detection.

Thus, the main result of this work is that neither single candidates for optical transients were detected under conditions of a low background near the zenith, nor candidate repeaters with large zenith angles in the background of an EAS were found. The absence of candidates for optical transients allows us to set limits on the frequency of occurrence of such events. Since the exposition for the season 2018--2019 is known (288~sr$ \cdot$hour), and no candidate was detected, then, taking into account the thresholds and signal duration that could be observed, the limitation on the flux of the sought transients can be formulated in the following way: for optical transients with an energy flux more than $10^{-4}$\,erg/cm$^2$/s and pulse duration more than 5\,ns the flow is less than $2\times10^{-3}$~events/sr/hour.

In conclusion, we note that the data of the HiSCORE array for the seasons 2019--2020 and 2020--2021 are awaiting of processing. At the same time, since the size of the array grows every year, a noticeable improvement in the background conditions for the search for optical transients in the HiSCORE data is expected.

The work was performed at the UNU ``Astrophysical Complex of MSU-ISU'' (agreement 13.UNU.21.0007). The work is supported by the Russian Science Foundation (grant no. 19-72-20067 (Section 3), the Russian Federation Ministry of Science and High Education (projects FZZE-2020-0017,  FZZE-2020-0024). The work was (partially) performed as part of the government contract of the SAO RAS approved by the Ministry of Science and Higher Education of the Russian Federation. G. Beskin acknowledges the support of Ministry of Science and Higher Education of the Russian Federation under the grant no. 075-15-2020-780 (N13.1902.21.0039).


\pagebreak

\end{document}